\documentclass[aps,prb,preprint,superscriptaddress,showpacs]{revtex4}
\usepackage{amsmath,amssymb,amsfonts}
\usepackage{graphicx}
\DeclareGraphicsExtensions{.jpg,.pdf,.png,.eps}
\newcommand{\mW} {$\mu \Omega$ cm }
\newcommand{\mWf} {$\mu \Omega$ cm}
\newcommand{\CVS} {CeVSb$_{3}$ }
\newcommand{\CVSf} {CeVSb$_{3}$}
\newcommand{\Tc} {$T_{C}$ }
\newcommand{\Tcf} {$T_{C}$}

\begin{document}
\title{Phase diagram of \CVS under pressure and its dependence on pressure conditions}
\author{E. Colombier}
\affiliation{Ames Laboratory, Iowa State University, Ames, Iowa 50011, USA}
\author{G. Knebel}
\affiliation{SPSMS, UMR-E CEA / UJF-Grenoble 1, INAC, Grenoble, F-38054, France}
\author{B. Salce}
\affiliation{SPSMS, UMR-E CEA / UJF-Grenoble 1, INAC, Grenoble, F-38054, France}
\author{E. D. Mun}
\altaffiliation[Currently at ]{NHMFL-Los Alamos}
\affiliation{Ames Laboratory, Iowa State University, Ames, Iowa 50011, USA}
\affiliation{Department of Physics and Astronomy, Iowa State University, Ames, IA 50011, USA}
\author{X. Lin}
\affiliation{Department of Physics and Astronomy, Iowa State University, Ames, IA 50011, USA}
\author{S. L. Bud'ko}
\affiliation{Ames Laboratory, Iowa State University, Ames, Iowa 50011, USA}
\affiliation{Department of Physics and Astronomy, Iowa State University, Ames, IA 50011, USA}
\author{P. C. Canfield}
\affiliation{Ames Laboratory, Iowa State University, Ames, Iowa 50011, USA}
\affiliation{Department of Physics and Astronomy, Iowa State University, Ames, IA 50011, USA}
\date{\today}
\begin{abstract}
We present temperature dependent resistivity and ac-calorimetry measurements of \CVS under pressure up to 8 GPa in a Bridgman anvil cell modified to use a liquid medium and in a diamond anvil cell using argon as a pressure medium, respectively. We observe an initial increase of the ferromagnetic transition temperature \Tc with pressures up to 4.5~GPa, followed by decrease of \Tc on further increase of pressure and finally its disappearance, in agreement with the Doniach model. We infer a ferromagnetic quantum critical point around 7~GPa under hydrostatic pressure conditions from the extrapolation to 0~K of \Tc and the maximum of the \textit{A} coefficient from low temperature fits of the resistivity  $\rho (T)=\rho _{0}+AT^{n}$. No superconductivity under pressure was observed down to 0.35~K for this compound. In addition, differences in the \Tcf\textit{(P)} behavior when a slight uniaxial component is present are noticed and discussed and correlated to choice of pressure medium.
\end{abstract}

\maketitle 
\section{Introduction}
\CVS is a member of the RVSb$_{3}$ (R=rare-earth) family, with an orthorhombic crystal structure. Two systematic studies \cite{Hartjes97,Sefat08} of this family showed interesting physical properties such as high anisotropy with a quasi two-dimensional crystal structure, and different types of magnetic order when the rare-earth is changed. Similarly complex properties were also observed in other binary and ternary rare-earth antimonide families, such as RSb$_{2}$ \cite{Budko98}, RCrSb$_{2}$ \cite{Leonard99} and RAgSb$_{2}$ \cite{Myers99}. 
\CVS is the only ferromagnetic compound from the RVSb$_{3}$ family, with \Tc around 4.6~K \cite{Sefat08}. It may be considered as a moderately heavy fermions system as its $\gamma$ value is found to be 162~mJ/mol K$ ^{2} $ below 2~K \cite{Sefat08}. Only a few studies involving this compound have been reported \cite{Hartjes97,Sefat08,Vannette08}.

Similar ferromagnetic, Ce-based compounds, such as CeNiSb$_{3}$ or CeAgSb$_{2}$ were studied under pressure by resistivity measurements \cite{Sidorov03,Nakashima03,Sidorov05Ce}, and revealed complex phase diagrams, with ferromagnetic transitions evolving into antiferromagnetic ones under pressure. Among the Ce-based ferromagnets studied under pressure to date, none have exhibited superconducting behavior.

The reported increase of \Tc for applied pressures up to 1~GPa \cite{Sefat08} motivated us to continue investigations on \CVS at higher pressures; ultimately it is reasonable to expect \Tc to pass through a local maximum value and then decrease \cite{Doniach77}. Ideally this would present a good opportunity to study possible quantum criticality in a Ce-based ferromagnet. We present here resistivity and ac-calorimetry measurements under pressure up to 8 GPa, in a Bridgman anvil cell modified to use a liquid medium and a diamond anvil cell, respectively. \Tc behaves as expected from the Doniach model \cite{Doniach77} with an initial increase with pressure up to a maximum above which a fast decrease and ultimately disappearance of \Tc is observed. The low temperature power law fits of the resistivity are in agreement with the disappearance of the magnetic transition at a quantum critical point.

We observed discrepancies in the \Tcf\textit{(P)} behavior between pressure cells using different pressure media, and attributed it to the differences of pressure conditions, with a slight uniaxial component existing along the cell axis in the Bridgman anvil cell.  Further measurements in the modified Bridgman cell with a different, more hydrostatic, pressure medium, confirmed this assumption. With differently oriented samples we have studied the \Tcf\textit{(P)} dependencies for different directions of the uniaxial component of pressure.

\section{Experimental details}

Single crystals of \CVS were grown out of antimony flux as detailed by Sefat \textit{et al.} \cite{Sefat08}. Resistivity and specific heat measurements were performed on these crystals up to 7.6 and 6.9~GPa, respectively. 

The resistivity samples were measured by a four probe method using the AC-transport option of a Quantum Design Physical Property Measurement System (PPMS) down to 1.8~K or a LakeShore 370 AC resistance bridge with a $ ^{3} $He cryostat down to 400~mK. Four, 12.5~$ \mu m $ diameter, gold wires were spot-welded to each polished and cut crystal with typical dimensions of 600 $\times$ 150 $\times$ 40~$\mu m^{3}$. Unless specified otherwise, the resistivity was measured along the \textit{c}-axis with the sample larger dimensions respectively along the \textit{c} and \textit{b}-axis. The measurement current was 1~mA and the frequency was 17~Hz. Before each sample was loaded into the pressure cell, the resistivity was measured at ambient pressure on a standard PPMS puck. A reproducible \Tc of 4.56~K was deduced from a sharp peak in the resistivity derivative, similarly to an averaged value of 4.6~K found previously \cite{Sefat08}.

Before performing studies under pressure, we measured several samples at ambient pressure with current flowing along each of the three crystallographic directions of the orthorhombic structure (at least two samples for each direction). We observed a good reproducibility in the resistivity behavior, although the uncertainties associated with measuring the relatively small sample dimensions lead to an error in the resistivity value at room temperature of up to 30-40~\% . Each sample's orientation was identified from the crystal's morphology, as was discussed by Sefat \textit{et al.} \cite{Sefat08} without any further X-ray Laue measurements. The reproducibility in resistivity from one sample to another was considered an indication that contributions from the other components of the resistivity were low or absent.

In addition, thermal expansion was measured at ambient pressure using a capacitive dilatometer constructed of OFHC copper, mounted in a Quantum Design PPMS instrument. A detailed description of the dilatometer is presented elsewhere \cite{Schmiedeshoff06}. The samples were lightly polished so as to have parallel surfaces approximately parallel to the different crystallographic axis directions. The dimensions range from 0.5~mm to a few mm. Measurements were performed on warming.  We define the thermal expansion coefficients as $ \alpha_{i}=\dfrac{1}{L_{i}}\dfrac{dL_{i}}{dT} $ with $L_{i}$ being one of the 3 sample's principle crystallographic orientations, and the volume thermal expansion coefficient $ \beta=\Sigma(\alpha_{i}) $.

Resistivity measurements under pressure were performed using a Bridgman cell modified to use with a liquid pressure medium \cite{Colombier07,Colombier09}, either a Fluorinert mixture (1:1 FC70:FC77) or 1:1 n-pentane:isopentane. When not specified, the medium used was 1:1 FC70:FC77. A piece of lead, to use as a manometer, and the sample were inserted in a pressure chamber of 1.4~mm inner diameter. The typical transition widths for lead were 15~mK and 40~mK, respectively for 1:1 n-pentane:isopentane and 1:1 FC70:FC77.  

Although, ideally, we would like pressure to be hydrostatic (i.e. isotropic), even with a medium that is a liquid at ambient conditions, once the media freezes at room temperature the application of pressure is expected to give rise to non-hydrostaticity that, as a first approximation, can be thought of as an application of hydrostatic pressure as well as a smaller uniaxial pressure. Given our cell geometry, if such a uniaxial component exists, it is anticipated to be in the direction perpendicular to the thin-disk-like sample space volume, i.e. along the cell axis. If we align the sample with one of its crystallographic axes along this direction, then we will say that "pressure is applied along this direction" to identify this potential uniaxial direction. For example, we use in the following the notation $ \rho_{c,P//a} $ to refer to the resistivity measured with current along the \textit{c}-axis and with pressure applied along the \textit{a}-axis of the sample.

Although the pressure is not purely hydrostatic, it is reproducible. Three samples were measured with similar pressure conditions (Bridgman cell filled with Fluorinert), current along the \textit{c} axis and pressure applied along the \textit{a} axis. The reproducibility of results was confirmed by similar \textit{T(P)} phase diagram data.

1:1 n-pentane:isopentane offers pressure conditions much closer to hydrostaticity, compared to 1:1 FC70:FC77 in the Bridgman cell pressure range, as it is known to freeze above 5~GPa at 300~K instead of below 1~GPa for the Fluorinert mixture \cite{Sidorov05}. However, it is more difficult to handle because of its high compressibility in the low pressure range \cite{Stella_bientot} (below 2~GPa) and because its boiling point is close to room temperature (28.5$^{\circ}$C for isopentane). Due to these difficulties, one of the resistivity data sets, in 1:1 n-pentane:isopentane media, was taken in a three wire configuration after the failure of one of the wires. The resulting three wires resistivity measurement gave limited quantitative information, but a sharp transition was still observable, and its derivative, (shown in figure \ref{Derivative}.a. below) looks very similar to those obtained from four wires measurements, once the data for the first pressure are scaled to ambient pressure.

The specific heat under pressure was measured in a diamond anvil cell \cite{Salce00,Demuer00} up to pressures of 7~GPa cooled down to 1.5~K using a $^{4}$He cryostat. The culet size of the anvils was 0.7~mm. The pressure, changed \textit{in-situ} at low temperatures \cite{Salce00}, was read using the ruby fluorescence method. Argon was chosen as a pressure medium. Albeit solidified at 1.4~GPa and 300~K, argon provides close to hydrostatic conditions due to its weak interatomic interactions (i.e a very soft solid). Three different pressure runs were performed. For one of these, two pressure cycles were realized by decreasing pressure in one step after a first run with increasing pressure. The ac-calorimetry method \cite{Demuer00} was used; a quasi-sinusoidal excitation was applied to the sample by a laser via a mechanical chopper. The temperature oscillations of the sample (inversely proportional to the specific heat) were measured with a Au/AuFe (0.07\%) thermocouple which was spot-welded on the sample. We estimated the amplitude of temperature oscillations of the sample $T_{ac} $ from the thermocouple voltage measured $V_{ac} $ and the thermoelectric power of the thermocouple $S_{th} $: $ T_{ac}= \vert V_{ac} \vert / S_{th} $. 

\section{Results}
\subsection{Ambient pressure}
\begin{figure}[!ht]
\begin{center}
\includegraphics[angle=0,width=120mm]{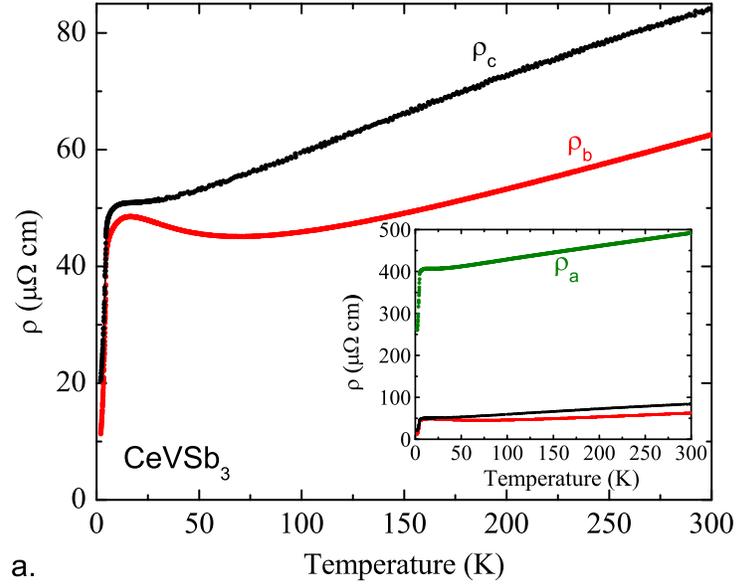}
\includegraphics[angle=0,width=120mm]{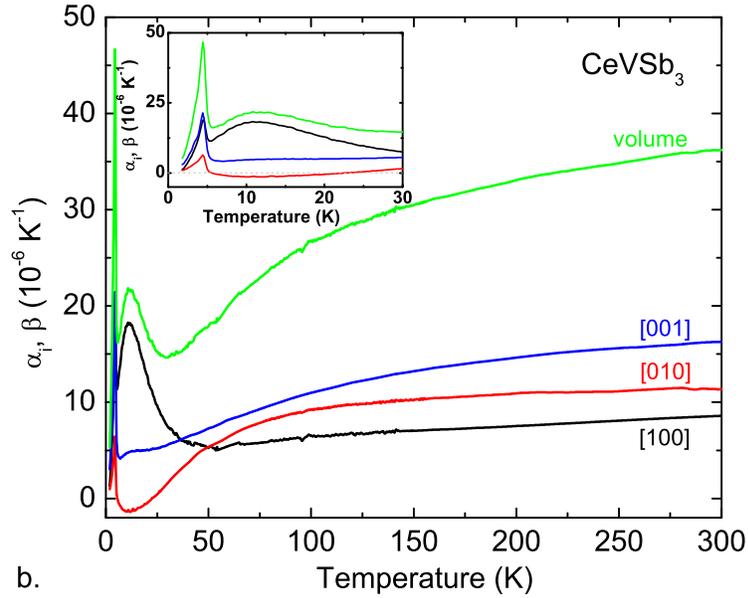}
\end{center}
\caption{(Color online) a. Resistivity, at ambient pressure, of \CVS along the \textit{b} and \textit{c} axis. The resistivity along the third direction is added to the two other in the inset. b. Anisotropic thermal expansion coefficients of \CVSf , inset shows expanded, low temperature range.}
\label{anisot}
\end{figure}

Our examination of \CVS (\textit{a}=13.172 \AA , \textit{b}=6.2419\AA , \textit{c}=6.0327 \AA) \cite{Sefat08} under pressure includes the study of the anisotropic properties of \CVS and in particular its sensitivity to slight uniaxial strains. In order to accomplish this, we first investigated the anisotropic resistivity with current \textit{i} flowing along the three crystalline directions of this compound at ambient pressure (figure \ref{anisot}.a). The resistivity ratios between 300~K and 2~K range from 2, when the current flows along the \textit{a}-axis, to 5.5 along the \textit{b}-axis. The results for current along the \textit{b} and \textit{c} axis are consistent with the study from Sefat \textit{et al.} \cite{Sefat08}, although resistivity values are lower in our measurements. 
A clear local maximum is observed at around 16~K for \textit{i} along the \textit{b}-axis, and is barely detected for \textit{i} along the \textit{c}-axis. A more striking anisotropy is the resistivity measured with current along the \textit{a}-axis, roughly 10 times higher than along the two other directions.  
The resistivity thus tends to be quasi two-dimensional, and in the following we study the resistivity along the \textit{b} or \textit{c} axis, depending on the geometry needed. 

In addition, a strong anisotropy in the thermal expansion coefficients is shown in figure \ref{anisot}.b. The broad local maximum of the volume thermal expansion coefficient around 10~K may be related to the Kondo temperature. We applied the Ehrenfest relation for second order phase transitions,
\begin{center}
$ \dfrac{dT_{C}}{dP_{i}}=\dfrac{V_{m}\Delta \alpha_{i}T_{C}}{\Delta C_{p}} $ ; $ \dfrac{dT_{C}}{dP}=\dfrac{V_{m}\Delta \beta T_{C}}{\Delta C_{p}} $ 
\end{center}
where $V_{m}$ is the molar volume, $ \Delta \alpha_{i} $ and $ \Delta \beta $ are respectively a change in the linear or volume thermal expansion coefficients at the phase transition, and $ \Delta C_{p} $ is a change in the specific heat \cite{Sefat08} at the phase transition. From this relation, we deduced a substantial uniaxial pressure dependence anisotropy for \Tcf , of 0.4 K/GPa, 0.2 K/GPa and 0.7 K/GPa when the pressure is respectively applied along the \textit{a}, \textit{b} and \textit{c} axes. The addition of these three components gives \textit{d}\Tcf \textit{/dP}=1.4~K/GPa, very close to the low pressure slope \textit{d}\Tcf \textit{/dP}=1.2~K/GPa found from the pressure temperature phase diagram (see figure \ref{PD} below).

\subsection{Resistivity under pressure}
The main interest of the present study is in the investigation of the evolution of \Tc under pressure. The modified Bridgman cell used with Fluorinert as a pressure medium is known to present a slight uniaxial component in addition to the expected isotropic pressure due in part to its low, room temperature freezing pressure, below 1~GPa. As an example, the iron arsenide superconductors, recently measured in this cell \cite{Colombier09,Colombier10}, are known to be sensitive to these uniaxial stresses which stabilize the superconducting phase. This superconducting phase is then observed in a broader pressure range of the phase diagram in presence of a uniaxial component of pressure. Since \CVS is an orthorhombic compound with clear anisotropy and some degree of electronic correlation, we decided to check its sensitivity to uniaxial component of pressure associated with non-hydrostaticity. The grown crystals are relatively large and mechanically sturdy making them easy to polish to three different geometries to allow for this study. (This is in contrast to iron arsenides, which were soft and easily exfoliated along their tetragonal, \textit{c}-axis.)

\begin{figure}[!ht]
\begin{center}
\includegraphics[angle=0,width=80mm]{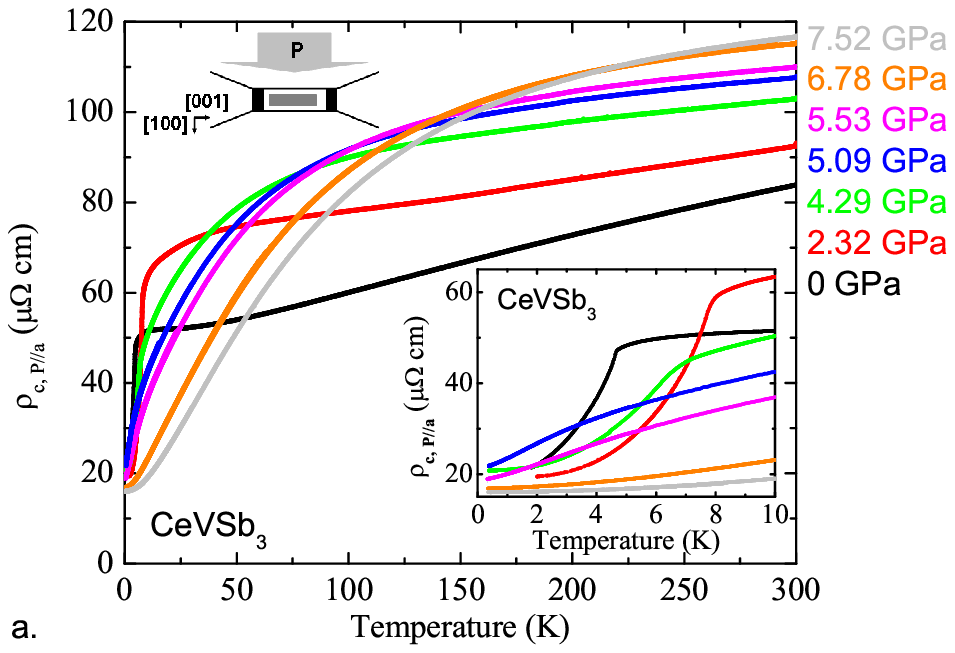}

\includegraphics[angle=0,width=80mm]{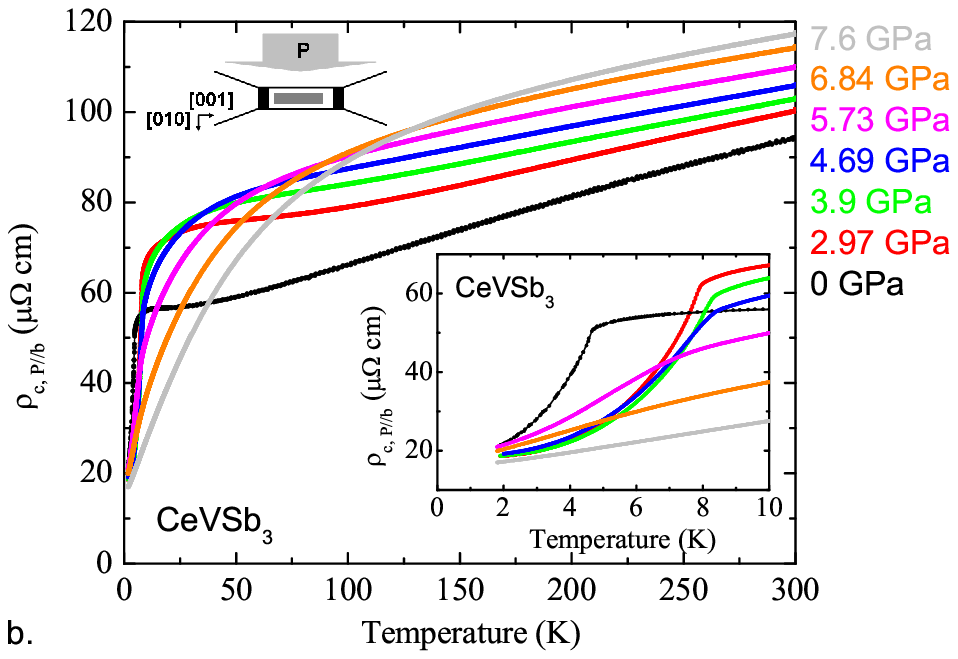}

\includegraphics[angle=0,width=80mm]{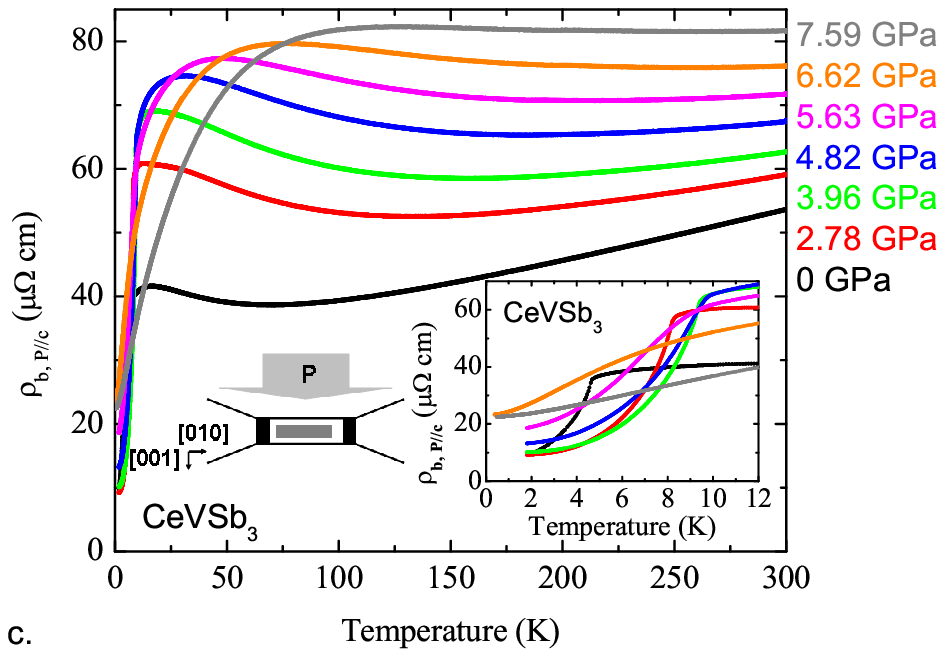}
\end{center}
\caption{(Color online) Resistivity measurement of \CVS under pressure. Sketches illustrate the sample orientations in the pressure cells. Insets: low temperature resistivity. a. with current along the \textit{c}-axis and pressure applied along the \textit{a}-axis.  b. with current along the \textit{c}-axis and pressure applied along the \textit{b}-axis. c. with current along the \textit{b}-axis and pressure applied along the \textit{c}-axis.}
\label{Resistivity}
\end{figure}

The temperature dependent resistivity data of \CVS measured with the pressure successively applied along the three crystallographic directions are shown in figure \ref{Resistivity}. We had to measure the resistivity along two different directions so as to fully investigate the response of the crystal to slight uniaxial stresses, but the evolution of the anisotropy of resistivity under hydrostatic pressure was not the main purpose of this work. 

In all cases, the resistivity above \Tc increases with pressure. \Tc itself initially increases with pressure, reaches a maximum value, and then decreases with pressure and finally disappears. The transition is sharp at ambient pressure and broadens progressively. It is difficult to distinguish it as \Tc drops towards 0~K. The resistivity curves presented in figures \ref{Resistivity}. a. and b. are obtained with the same current direction, but the transition temperature increase is slower in b (see \textit{T(P)} phase diagram in figure \ref{PD} below). This shows evidence for anisotropy of the pressure response of the crystals, as the reproducibility of results was checked for three pressure runs in similar conditions. For current along the \textit{b}-axis (figure \ref{Resistivity}.c), the local maximum observed at ambient pressure is still present under pressure; it progressively broadens as it is shifted up to higher temperatures. For each direction of applied pressure, there is a clear and consistent increase of $ \rho _{300K} $ over the measured pressure range. 

\begin{figure}[!ht]
\begin{center}
\includegraphics[angle=0,width=53mm]{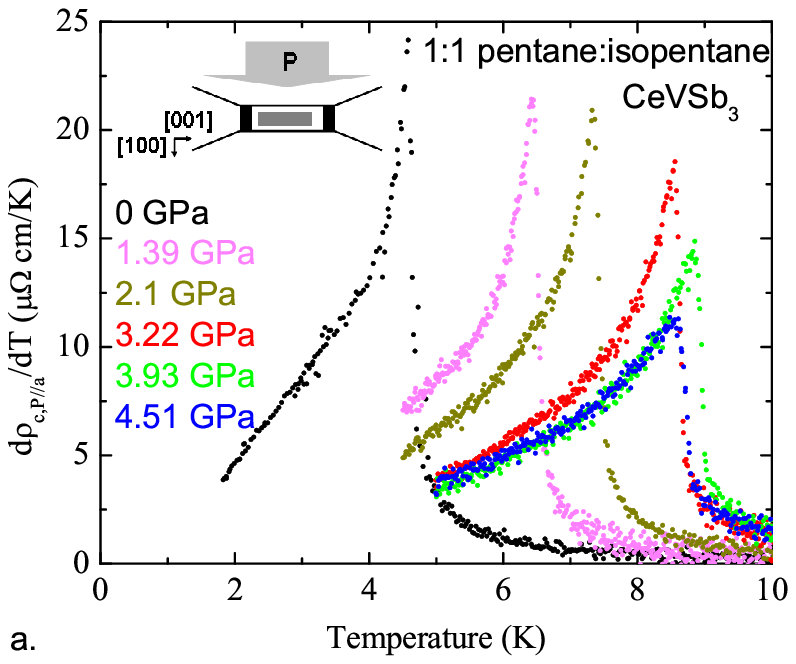}

\includegraphics[angle=0,width=53mm]{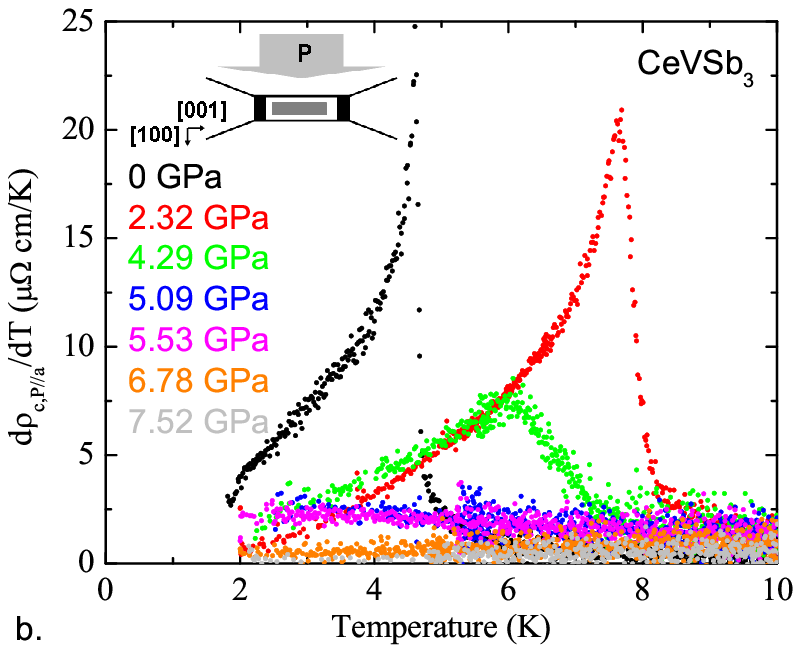}

\includegraphics[angle=0,width=53mm]{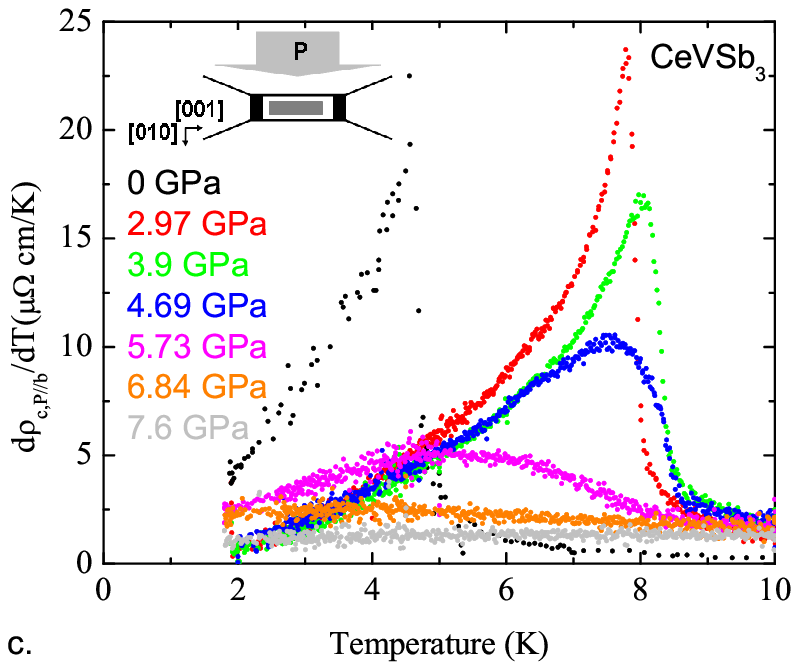}

\includegraphics[angle=0,width=53mm]{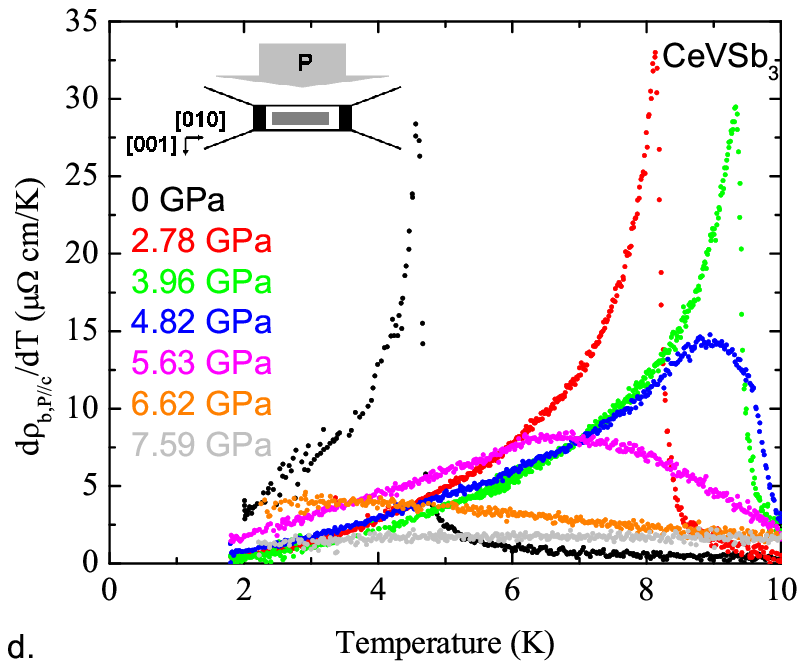}
\end{center}
\caption{(Color online) Resistivity derivative $ d\rho (T)/dT $ of \CVS under pressure. Sketches outline the sample orientations in the pressure cells. a. with current along the \textit{c}-axis and pressure applied along the \textit{a}-axis in a cell filled with 1:1 n-pentane:isopentane (in $\mu \Omega$ $cm /K  $ at 0~GPa and arbitrary units under pressure). b. with the same orientation, but filled with 1:1 FC70:FC77 c. with current along the \textit{c}-axis and pressure applied along the \textit{b}-axis with 1:1 FC70:FC77 as a pressure medium. d. with current along the \textit{b}-axis and pressure applied along the \textit{c}-axis with 1:1 FC70:FC77 as a pressure medium.}
\label{Derivative}
\end{figure}

The low temperature resistivity derivative data, $ d\rho (T)/dT $, are compared in figure \ref{Derivative}, for the three different cells configuration shown in figure \ref{Resistivity} as well as an additional cell filled with 1:1 n-pentane:isopentane. The influence of sample orientation on \Tc is even more obvious when the data are presented in this manner. The highest \Tc value is observed in fig. \ref{Derivative}.d, for the \textit{c}-axis of the sample aligned with the cell axis.  In the graphs a. and b., the samples' orientations are the same but two different pressure media are used: 1:1 n-pentane:isopentane and 1:1 FC70:FC77, respectively. We observe a strong dependence on pressure conditions. Whereas the feature remains sharp until the highest pressure of 4.5~GPa with 1:1 n-pentane:isopentane (figure \ref{Derivative}.a), it is already broadened significantly at a similar pressure in Fluorinert (figure \ref{Derivative}.b), and \Tc is much lower. In the experiment with Fluorinert, the transition temperature broadens significantly for pressures above 4~GPa. 

\subsection{Phase diagram}
\begin{figure}[!ht]
\begin{center}
\includegraphics[angle=0,width=90mm]{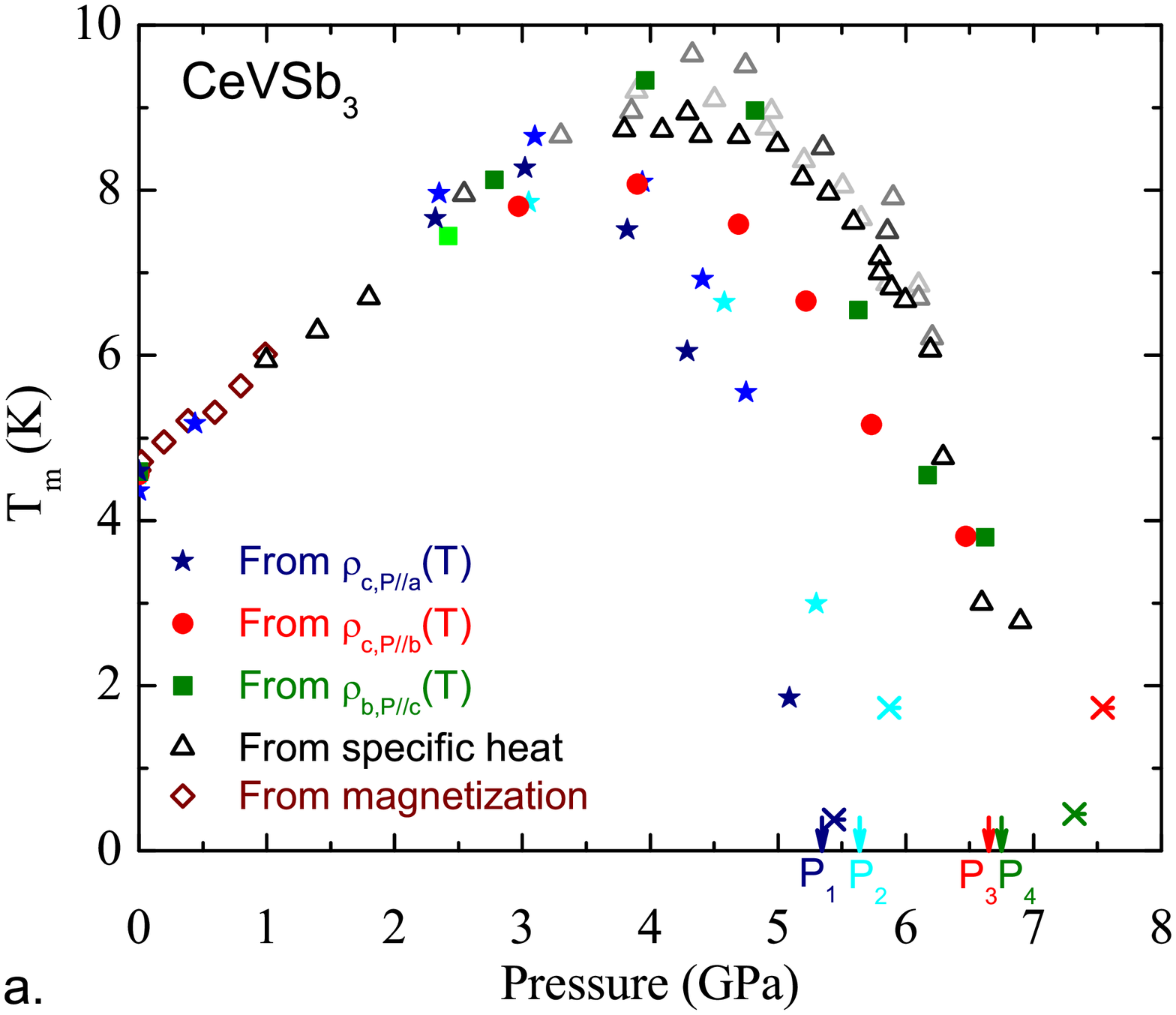}
\includegraphics[angle=0,width=90mm]{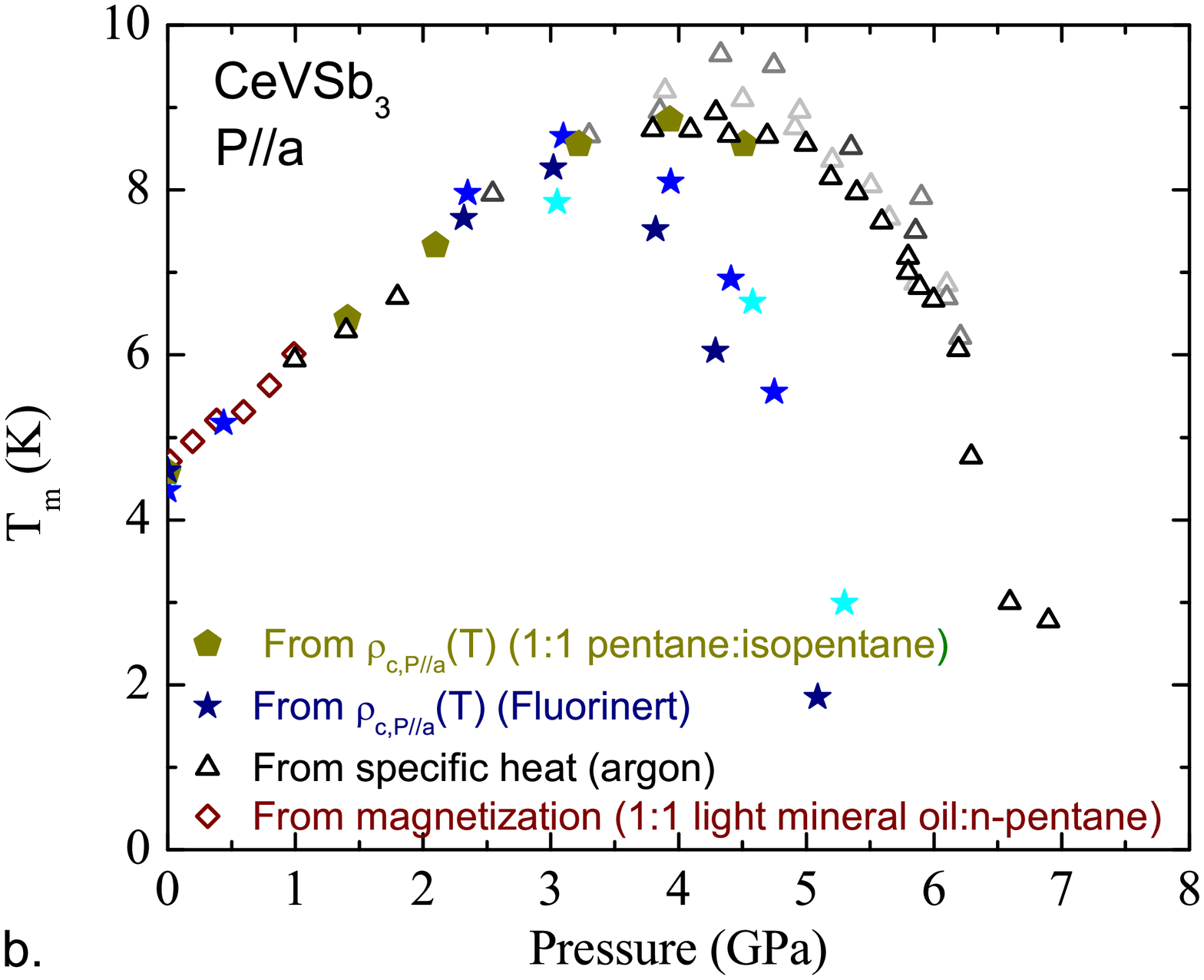}
\end{center}
\caption{(Color online) \textit{T(P)} phase diagram of \CVSf . We added data from magnetization measurements performed in a piston-cylinder cell \cite{Sefat08}. For a given symbol, experimental conditions were similar, and the different colors refer to different runs. a. Comparison between the different axis orientations in the Bridgman anvil cell. Diamond anvil cell $C _{P} $ data points are also shown. Crosses are the lowest measured temperature for the lowest pressure for which no phase transition could be detected. $P_{1}$, $P_{2}$, $P_{3}$ and $P_{4}$ are given by downwards arrows and refer to the critical pressures estimated from figure \ref{LTfit}. b. Comparison between the different pressure media used.}
\label{PD}
\end{figure}

Figure \ref{PD} shows the phase diagrams obtained from several runs with different pressure conditions and different crystal orientations in the modified Bridgman cell, together with data points inferred from the piston-cylinder cell magnetization data and the diamond anvil cell specific heat data. 
For each pressure run, we observe a similar dome-shaped phase diagram. However, the data from runs with different media and orientations are somewhat scattered. Whereas all curves overlap below at least 2~GPa, differences in \Tcf \textit{(P)} are observed at higher pressures. 

We observe obvious differences between the 3 crystal orientations measured in the modified Bridgman cell, figure \ref{PD}.a. The maximum values of \Tc range from 7.9~K to 9.3~K and the corresponding pressures from 3.2~GPa to 4.2~GPa. More importantly, the critical pressure, the pressure at which the \textit{T(P)} curve extrapolates to zero, ranges from roughly 5.5 to 7~GPa. As this behavior is reproducible within 0.3~GPa for three different runs when the cell axis coincides with the \textit{a} crystallographic axis, we assume the differences seen come from an anisotropic response to the slight uniaxial component present in the modified Bridgman cell. This strong anisotropy and particular sensitivity to uniaxial component of pressure along the \textit{a} crystallographic axis is confirmed when we use a more hydrostatic pressure medium. Two runs in the modified Bridgman cell with the same crystal orientation (which appears to be the most sensitive to uniaxial pressure) are shown figure \ref{PD}.b, one with Fluorinert and one with 1:1 n-pentane:isopentane as a pressure medium. Here again we observe differences between the two runs in the maximum value of \Tcf , its corresponding pressure, and the critical pressure. 1:1 n-pentane:isopentane is known to freeze at room temperature above 5~GPa and Fluorinert freezes below 1~GPa \cite{Sidorov05}. This means that, contrary to the other medium, 1:1 n-pentane:isopentane was always liquid at room temperature in this experiment, and so was much closer to hydrostaticity. (A conclusion established by the superconducting transitions widths of the lead manometers, given in the experimental details section.) The basic agreement between the 1:1 n-pentane:isopentane data and the $C_{p} $ data taken in argon (discussed below) and the higher pressure deviation of the Fluorinert data from this manifold is further evidence that the discrepancies in the phase diagram can be attributed to an anisotropic sensitivity of the sample to a uniaxial component of pressure. Keeping in mind the strong sensitivity of \CVS to pressure conditions, we try to be very cautious about the impact of pressure conditions to our results.

\begin{figure}[!ht]
\begin{center}
\includegraphics[angle=0,width=90mm]{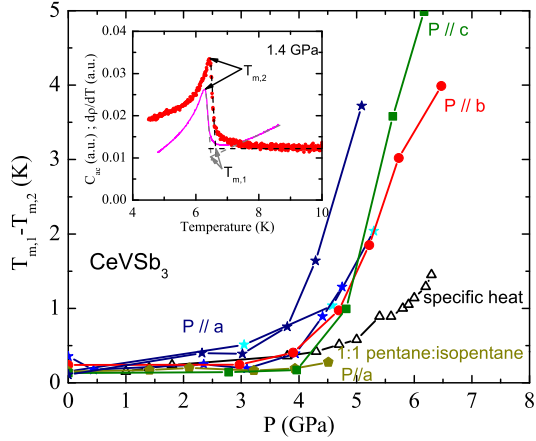}
\end{center}
\caption{(Color online) Evolution with pressure of the difference in temperature between two criteria for the magnetic transition. Several curves are shown for different resistivity measurement conditions and one for the specific heat. Three data sets are shown for \textit{P//a} (stars). The inset shows the definition for these two criteria, for the resistivity derivative and the specific heat.}
\label{Delta_TCurie}
\end{figure}

To estimate the evolution of the samples' sensitivity to pressure conditions, we checked the broadening of the magnetic transition. The lead, as a soft material, is not very sensitive to deviations to hydrostaticity and the broadening of the superconducting transition is modest \cite{Colombier09}. The transition broadening of \CVS would indeed be a more efficient clue as long as we are able to estimate also the effects intrinsic to the magnetism. We estimated in figure \ref{Delta_TCurie} the broadening of the transition by comparing two different criteria for \Tcf : the maximum of the peak in the $ d\rho /dT$ derivative and the onset of this peak from two asymptotes (as shown by dashed lines in the inset). By comparing cells measured in different pressure conditions, we get a good sense of pressure effect versus intrinsic properties of the compound. As $ d\rho /dT$ appears to be very similar to the heat capacity feature around the transition, similar criteria were applied to $C_{P}(T)$ data.

At ambient pressure, the difference between both criteria is around 200~mK in resistivity and it increases only slightly up to around 400~mK at 3~GPa. Above 3 or 4~GPa, the transition broadens strongly, to above 4~K in the modified Bridgman cell filled with Fluorinert. The broadening observed with the 1:1 n-pentane:isopentane set of measurements is at least a factor of two smaller, compared to Fluorinert, with only a slight increase at the highest pressure. We did not reach pressures above 4.5~GPa, because of the low compressibility of this medium and a pressure medium leak observed at the lowest pressure. Whereas the effect of non-hydrostatic conditions on the transition broadening is obvious, some degree of broadening may be an intrinsic property of the material once \Tc starts to decrease at high pressures, as it generally seems to occur at pressures above the maximum of the magnetic transition temperature.

To further our investigation of the influence of pressure non-hydrostaticity, a noble-gas as a pressure medium was used to provide a near hydrostatic reference. Even when it is solid, its low interatomic interactions indeed allow excellent pressure conditions, as can be seen in figure \ref{Delta_TCurie}. This experiment entailed the measurement of specific heat in a diamond anvil cell with argon as a pressure medium, and is described below.

\subsection{ac-calorimetry}
\begin{figure}[!ht]
\begin{center}
\includegraphics[angle=0,width=90mm]{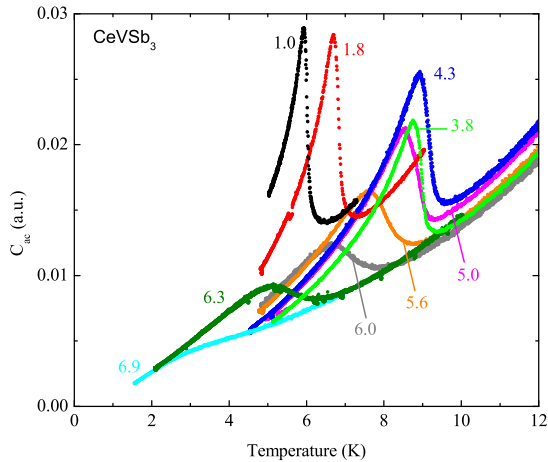}
\end{center}
\caption{(Color online) Specific heat  of \CVS (in arbitrary units) under pressure, measured in the diamond anvil cell.  Pressures are given in GPa.}
\label{Cp}
\end{figure}

In figure \ref{Cp}, we present the temperature dependent specific heat curves of \CVS obtained from one of the three pressure runs.
The transition at the lowest pressures is sharp with a shape similar to the ambient pressure measurement \cite{Sefat08}. The 1.0~GPa \Tc value inferred from the data presented in figure \ref{Cp} is in good agreement with that inferred from the magnetization data at 1.0~GPa \cite{Sefat08}. \Tc progressively increases with pressure until 4.3 GPa and then decreases. The transition progressively broadens, and its amplitude also seems to decrease, although the background and the signal amplitude might be a little different from one measurement to another. Whereas at pressures of 6.0 and 6.3~GPa, a feature is still clearly seen, we can just barely resolve a broad bump in the 6.9~GPa data. We can see in figure \ref{Delta_TCurie} the stronger broadening of the transition when \Tc decreases, similarly to what is found in the Bridgman cell resistivity data. However, as the pressure conditions are closer to hydrostaticity compared to a Bridgman cell filled with Fluorinert, the transition is twice as sharp, in particular at the highest pressures as seen in figure \ref{Delta_TCurie}. The agreement is much better when the pressure medium is 1:1 n-pentane:isopentane with sharp transitions in both cases, in the overlapping pressure range.

The \Tcf \textit{(P)} values obtained are also shown in the phase diagrams in figure \ref{PD}. The diamond anvil cell can be considered as the reference since the pressure conditions are presumed to be the closest to hydrostaticity. We observe an increase of \Tc from 4.6~K to as high as 9.7~K when the pressure increases from 0~GPa to 4.3~GPa. It then decreases and we expect to have a critical pressure around 7-7.5~GPa. From one run to another, only differences in maximum \Tc are noticed, and they are below 1~K. These differences may also be linked to pressure conditions. Light and medium gray triangles in figure \ref{PD} respectively refer to the first and the second pressure increase realized with the same diamond anvil cell in specific heat. We observe a slightly lower maximum \Tcf , around 0.5~K, for the second run, when the sample may be more strained. These differences between runs even with a noble gas as a pressure medium emphasize here again the extreme sensitivity of \CVS to pressure conditions. The diamond anvils cell axis (along which the load is applied) is coincident with the \textit{a}-crystallographic axis, and the phase diagram is moved up to higher pressures and temperatures, compared to the Bridgman cell measurement using the same sample orientation. Compared to 1:1 n-pentane:isopentane, difference are more subtle, and seem to mainly consist in a lower value of maximum \Tc in the Bridgman cell. 

\subsection{Quantum criticality}
\begin{figure}[!ht]
\begin{center}
\includegraphics[angle=0,width=60mm]{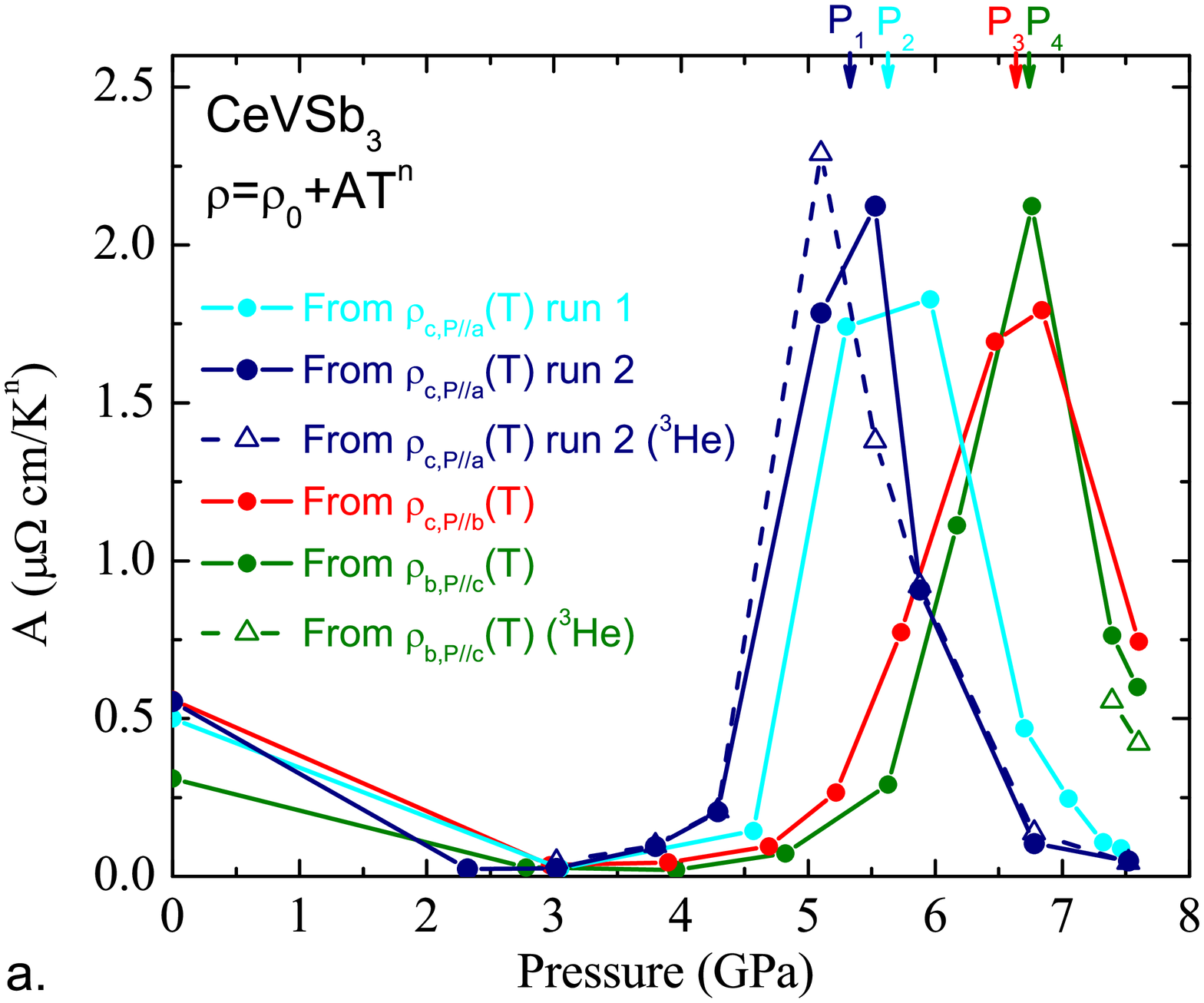}

\includegraphics[angle=0,width=60mm]{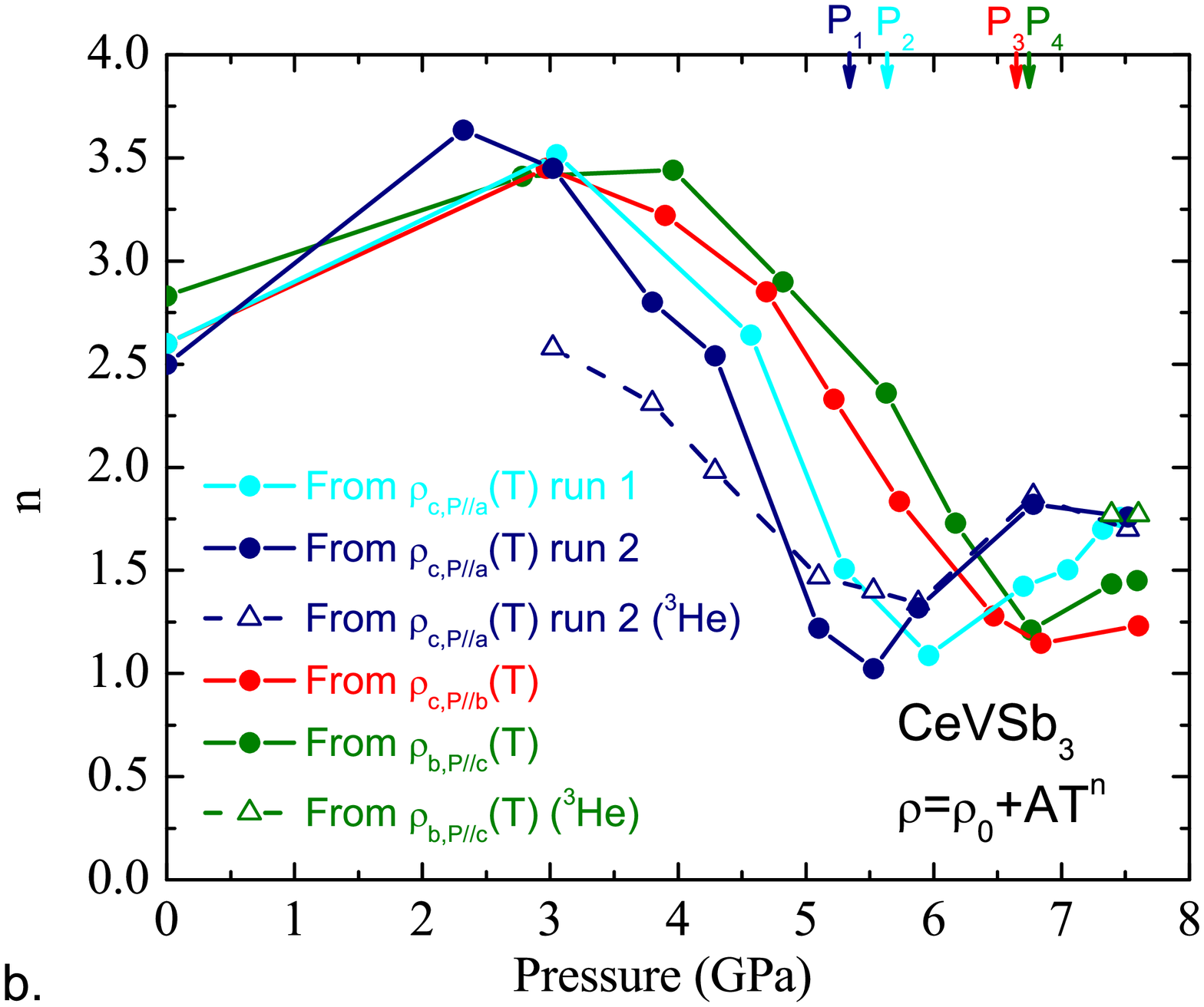}

\includegraphics[angle=0,width=60mm]{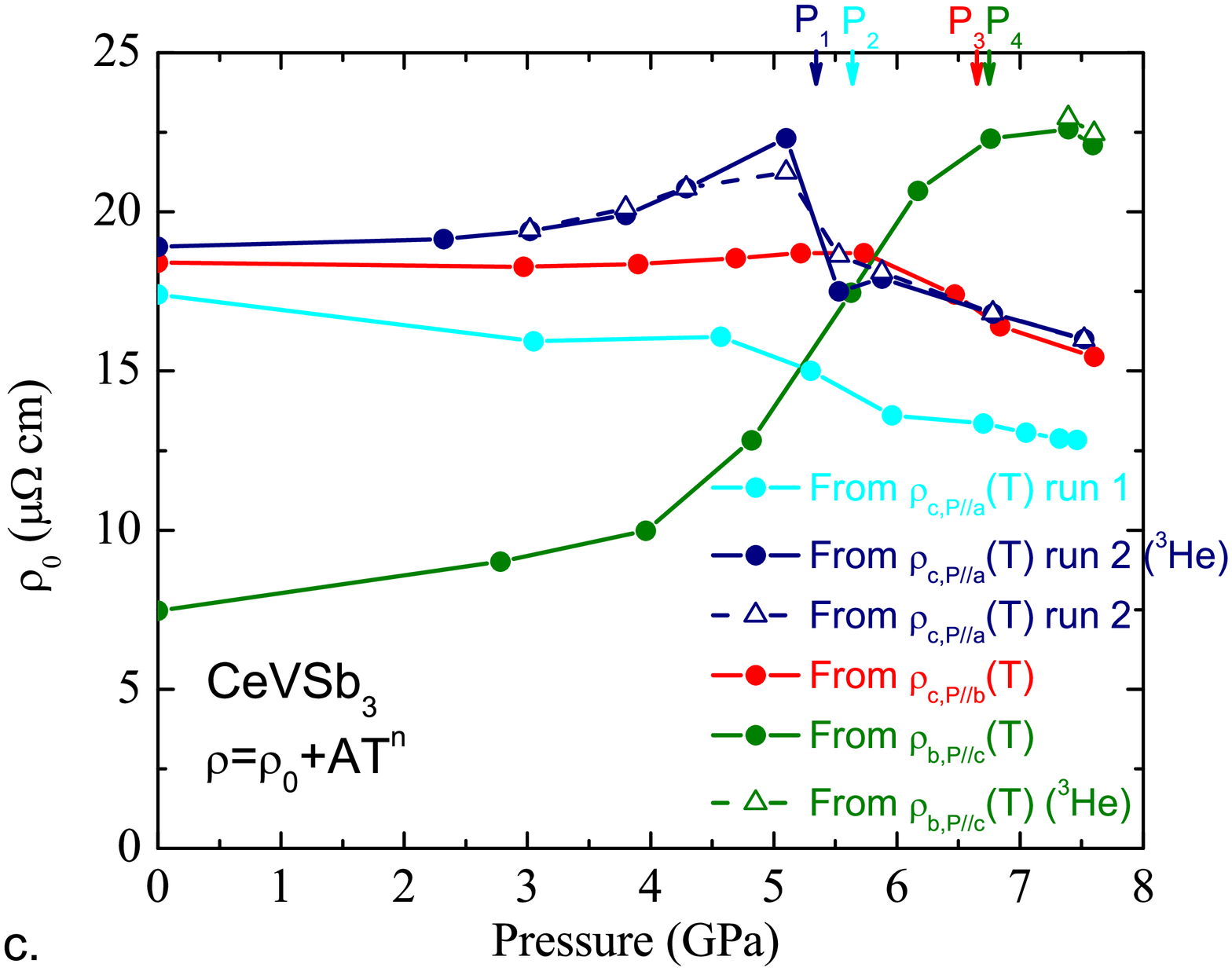}
\end{center}
\caption{(Color online) Pressure dependence of the parameters obtained from a low temperature fit $ \rho (T)= \rho _{0} + A T^{n} $. The arrows labeled $P_{1} $, $P_{2} $, $P_{3} $ and $P_{4} $ are the estimated critical pressures (see text). The triangles  refer to fits down to $ ^{3} $He temperatures. The colors are chosen the same as in figure \ref{PD}.  a. \textit{A} coefficient. b. Temperature exponent, \textit{n}. c. Residual resistivity, $\rho _{0}$. }
\label{LTfit}
\end{figure}

Whereas differences in the pressure dependence are noticed for the several pressure runs shown in the phase diagram figure \ref{PD}, the general behavior and in particular the way the magnetic transition is suppressed are similar. As our main interest is to determine if we observe a quantum critical point, we performed further low temperature measurements in a $ ^{3} $He cryostat. From these measurements we made low temperature resistivity fits using the equation: $ \rho (T)=\rho _{0}+AT^{n}$, where either (i) \textit{n} equals to 2 or (ii) \textit{n} was treated as a free fitting parameter. As measurements in a PPMS down to $ ^{4} $He temperatures are much more convenient, we performed only a few measurements down to 0.35~K, so as to check that the fits down to 1.8~K gave qualitatively similar results. Two  $ ^{3} $He measurements were performed above P$ _{c} $ when the pressure is applied along the \textit{c}-axis of the crystal and another whole set of measurements was made for \textit{P} $>$ 3~GPa with pressure applied along the \textit{a}-axis. Figure \ref{LTfit} presents fit data from when \textit{n} was left as a free parameter. We determined the temperature range of the fit either by a progressive increase of the maximum fit temperature, or by checking the linear behavior of the law $ Log(\rho (T) )=Log(\rho _{0} )+nLog(A)Log(T) $, when $\rho _{0}$ was slightly modified. The temperature ranges and fit results obtained from both methods were in good agreement, the maximum fit temperature being around 3.5-4~K for the measurements in a $ ^{4} $He cryostat. The results of fits performed in the $ ^{3} $He and in the $ ^{4} $He cryostat are in a good qualitative agreement but the parameters values (specifically \textit{n}) can differ of as much as 40\% around the critical pressure (where the magnetic transition disappears).

\begin{figure}[!ht]
\begin{center}
\includegraphics[angle=0,width=90mm]{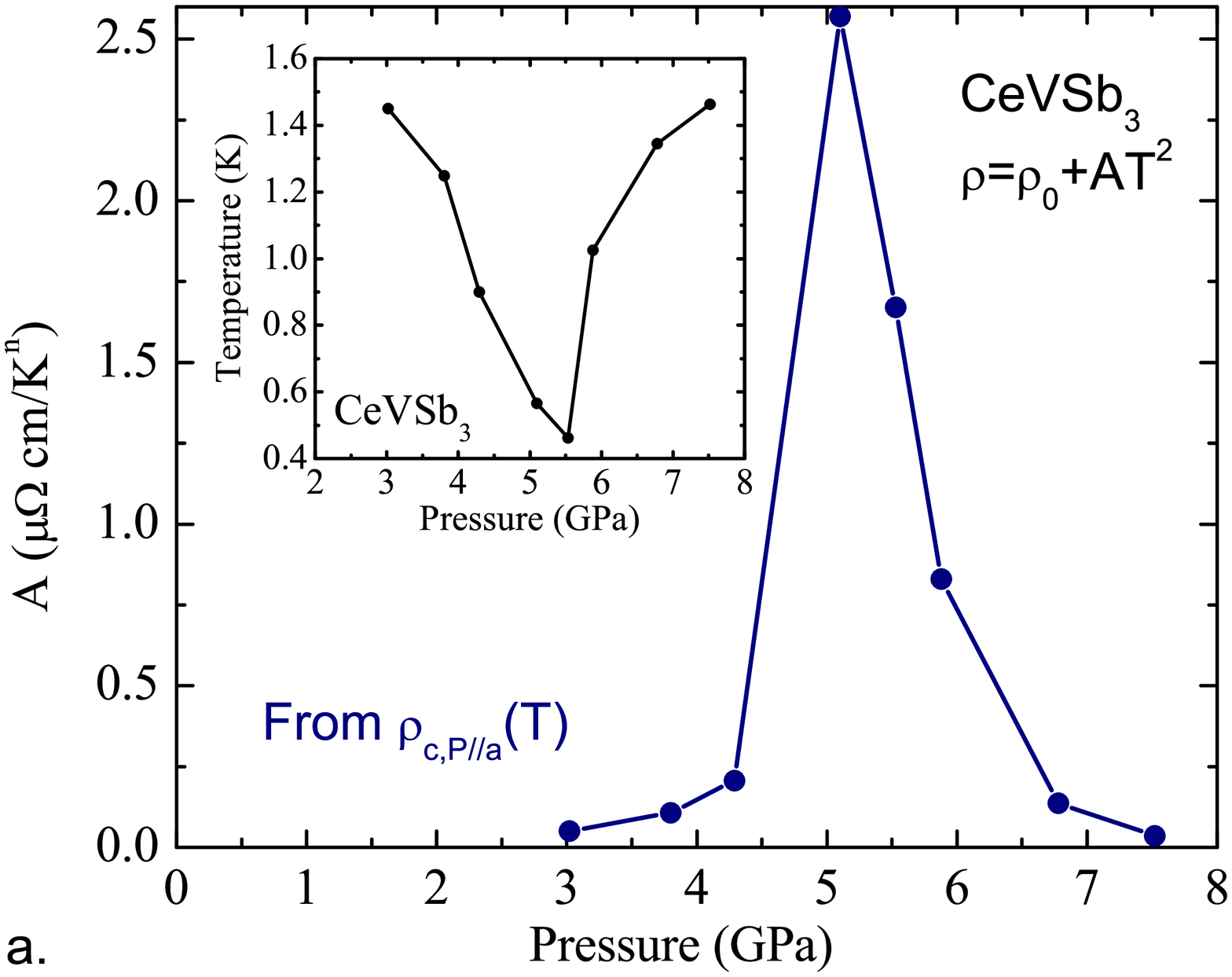}
\includegraphics[angle=0,width=90mm]{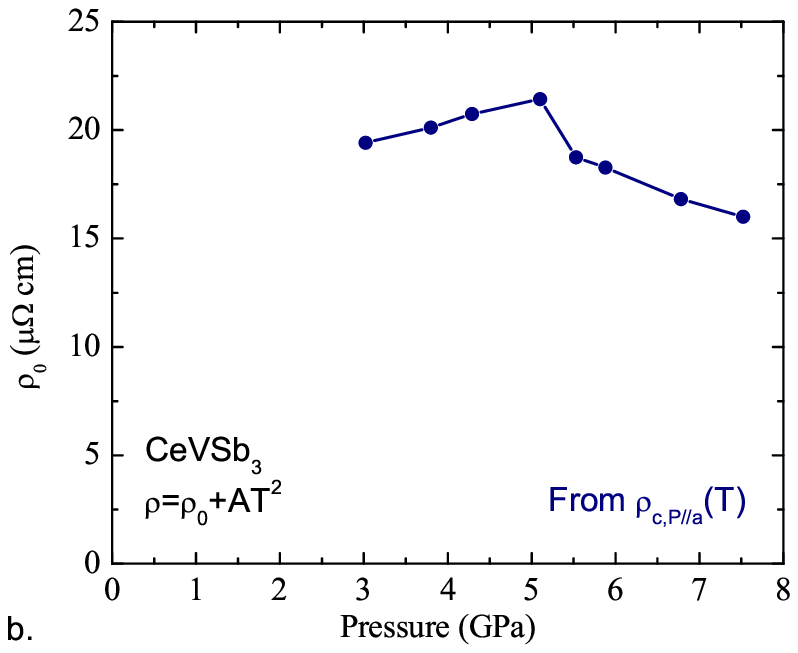}
\end{center}
\caption{(Color online) Pressure dependence of the parameters obtained from a low temperature fit $ \rho (T)= \rho _{0} + A T^{2} $ from $^{3}$He data of $ \rho_{c,P//a} $.    a. \textit{A} coefficient. The inset shows the maximum temperature where this fit applies. b. Residual resistivity, $\rho _{0}$. }
\label{LTfitT2}
\end{figure}

The general behavior of \textit{A} as well as \textit{n}, shown in figures \ref{LTfit}.a and \ref{LTfit}.b, is similar but shifted in pressure for the three orientations of the sample axes with respect to the cell axis. The \textit{A} parameter presents a strong peak around the pressure where the magnetic transition disappears. Roughly at the same pressure, a local minimum of the \textit{n} parameter can be observed. We estimated the critical pressures (labeled $P _{1} $, $P_{2} $, $P_{3} $ and $P_{4} $ in each panel of figure \ref{LTfit}) from the pressure average of the estimated local maximum of \textit{A} and minimum of \textit{n}. 
The critical pressures obtained this way with fits down to 1.8~K (cf figure \ref{LTfit}) were $P_{1} \approx$5.3~GPa (+/- 0.2~GPa), $P _{2} \approx$5.6~GPa (+/- 0.3~GPa), $P _{3} \approx$6.7~GPa(+/- 0.5~GPa) and $ P_{4} \approx$6.8~GPa(+/- 0.5~GPa). The error is due to the data spacing and the difference when $P_{c}$ is estimated from a fit of the \textit{T(P)} phase diagram \ref{PD}. 

$\rho_{0}$ behaves similarly when the cell axis is along the \textit{a} or \textit{b} crystallographic axis, with a slight increase around the pressure where \Tc disappears and a stronger decrease above. When the cell axis is along the \textit{c} crystallographic axis, the behavior is different, with a continuous increase which is faster in $ \sim $ 5-7~GPa range of pressures, once \Tc decreases. For this orientation, the current is along the \textit{b}-axis, instead of \textit{c}, which may have a strong influence of the pressure dependence of $\rho_{0}$. The \textit{RRR} for $ \rho _{b,P//c} $ decreases from 7.2 at 0~GPa to 3.5 close to the critical pressure. This is in contrast with the \textit{RRR} in the two other directions which monotonically increases from 4-5 at 0~GPa to nearly 8 above 7.5~GPa. 

Given the essentially complete $ \rho_{c,P//a} $ data set from our $ ^{3}$He run we can also try forcing the temperature exponent to be exactly equal to two at the lowest temperatures.  Figure \ref{LTfitT2} presents the pressure dependence of \textit{A} and $\rho_{0}$ as well as the temperature range over which the $T^{2} $ fit to the data could be made.  These results are consistent with those presented in Figure \ref{LTfit}:  there is a divergence in \textit{A} near 5.1~GPa and the temperature range that the data can be fit with a quadratic pressure dependence drops below our minimum temperature between 5 and 5.5~GPa.  

\section{Discussion}

\subsection{Anisotropy}
RVSb$ _{3} $ materials respond to chemical and applied pressure anisotropically. The lattice parameter decrease of RVSb$_{3}$ is anisotropic when R goes from La to Dy. Sefat \textit{et al.} \cite{Sefat08} found a decrease from  0.9\% to 5.4\% along the \textit{b} and \textit{a} axis, respectively. The thermal expansion of \CVS at ambient pressure (figure \ref{anisot}.b) is also clearly anisotropic and we deduced, from the Ehrenfest relation, a uniaxial pressure dependence anisotropy for \Tcf .

These observations motivated us to take advantage of the deviations from hydrostaticity in the modified Bridgman cell, to measure our samples with a slight additional uniaxial pressure component, successively applied along each of the three crystallographic axes. 
We already observed from figure \ref{PD} and \ref{LTfit} that the critical pressure is different depending on the lattice direction along which the pressure is applied. This difference is significantly larger than any cell-to-cell variation. When the uniaxial stresses are applied along a stiffer axis, the crystal may be subject to smaller distortions and a higher pressure would be needed to suppress the magnetic transition. From this picture, the \textit{c}-axis can be considered as the least sensitive to the Bridgman cell uniaxial component and give results closest to the ones obtained in the more hydrostatic diamond anvil cell. 

\begin{figure}[!ht]
\begin{center}
\includegraphics[angle=0,width=90mm]{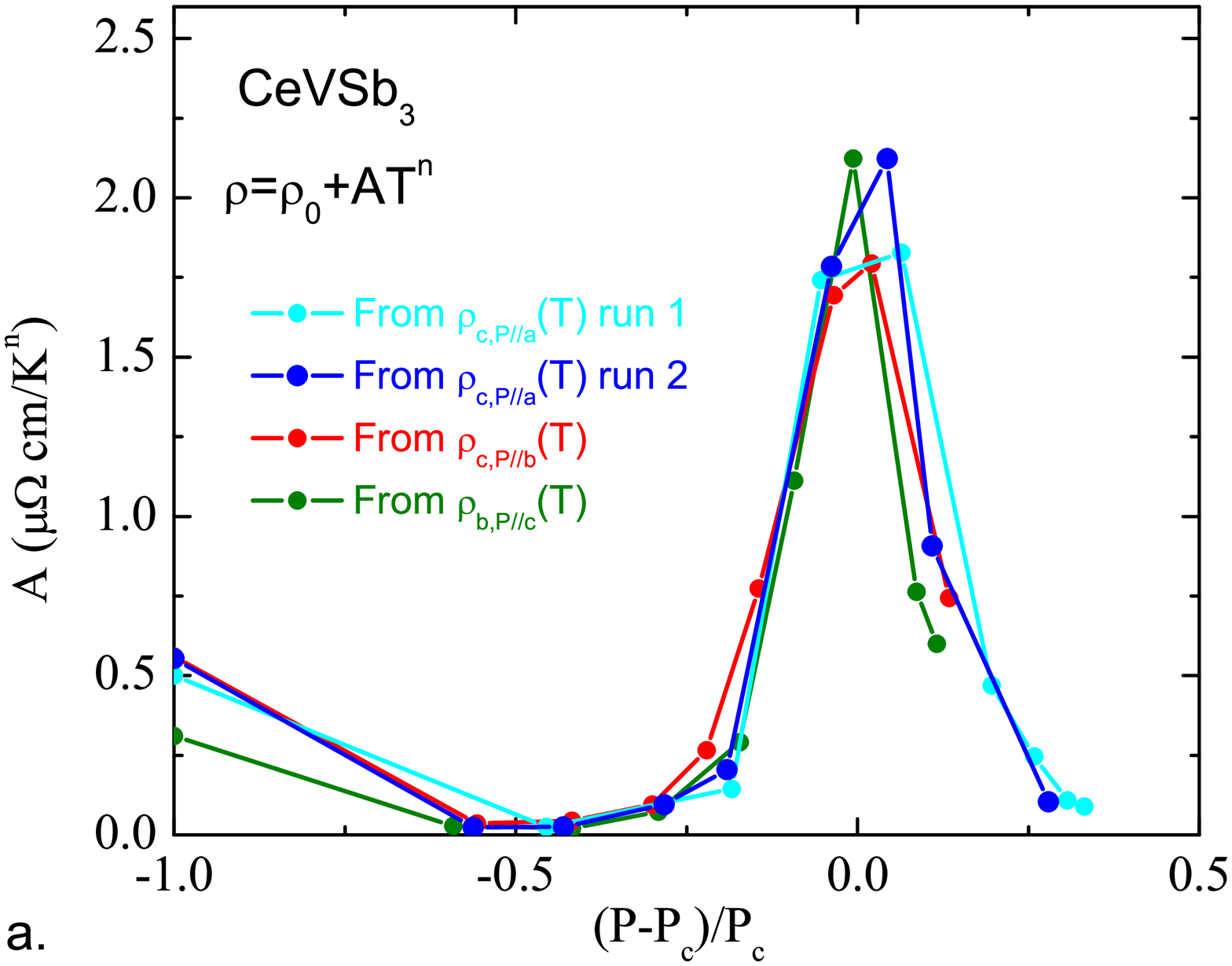}
\includegraphics[angle=0,width=90mm]{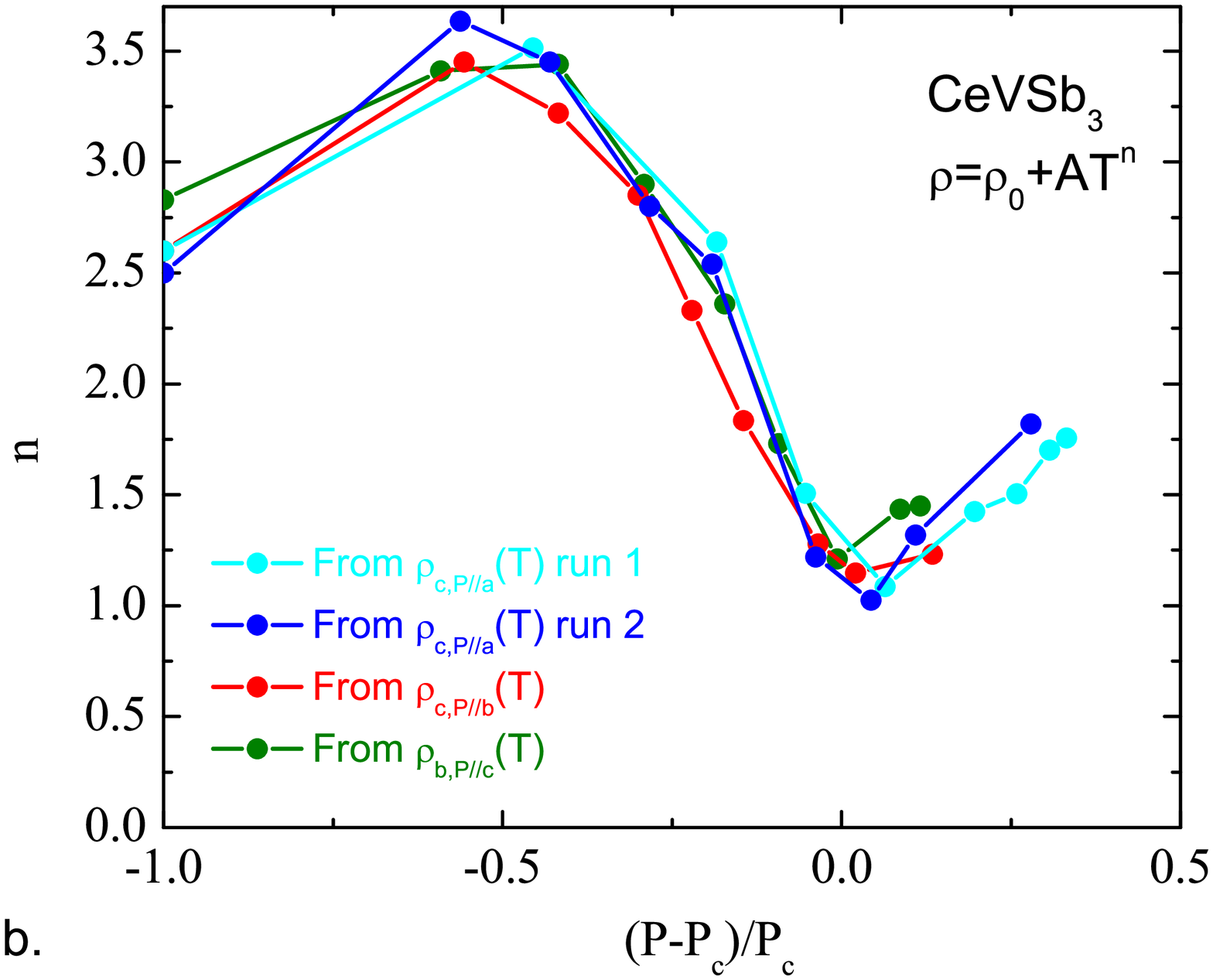}
\end{center}
\caption{(Color online) Dependence on a scaled pressure of the parameters obtained from a low temperature fit $ \rho (T)= \rho _{0} + A T^{n} $ ($T \geq $1.8~K). The critical pressure values, $ P_{c} $, used here were given in the results section from the low temperature fits of the resistivity. a. \textit{A} coefficient. b. Temperature exponent, \textit{n}.}
\label{LTfit_param}
\end{figure}

Whereas the uniaxial pressure dependence anisotropy deduced from the Ehrenfest relation gave us a clue to measure our samples with the pressure applied along several different crystallographic axis, the predicted anisotropy was not retrieved from our measurements at the lower pressures. We do not observe any deviations between the \Tcf \textit{(P)} curves below 3~GPa. This might be due to relatively good hydrostaticity in this pressure range.

It is interesting to notice that whereas deviations in hydrostaticity tend to modify the pressure dependence of \Tcf , the low temperature functional dependence of the resistivity appears to be similar at comparable distances from the critical point. Indeed if we define an effective pressure parameter as $ \dfrac{P-P_{c}}{P_{c}} $ (with $ P_{c} $, critical pressure determined above) we can plot all of the \textit{A} and \textit{n} data on this universal scale (Figure \ref{LTfit_param}). The fact that both the \textit{A} and \textit{n} data fall onto common manifolds indicates that the quantum critical behavior is inherent to the system and only depends upon the distance from the critical value of the tuning parameter, pressure in this case. This result implies that slight uniaxial components of pressure may be utilized to further tune the criticality without fundamentally changing the underlying physics.

\subsection{Phase diagram}
When pressure is applied, the magnetic ordering temperature first increases, passing through a maximum before decreasing at a faster rate. No magnetic transition is observed for pressures above 7~GPa. This behavior is consistent with what we expect from the competition between the Kondo effect and the Ruderman-Kittel-Kasuya-Yosida interaction. The phase diagram (figure \ref{PD}) has a shape in very good agreement with the Doniach model \cite{Doniach77}.

A goal of this study was to determine the presence of a possible quantum critical point. Although the suppression of \Tc seems continuous, we can not clearly follow the transition for $T < $ 1.5~K. Even between 1.5~K and at least 4~K, in the modified Bridgman cell as well as in the diamond anvils cell, the peak used to infer the transition temperature is broad and its amplitude is small. This broader transition and lower amplitude may be an additional evidence of the progressive weakening of the magnetic transition once the pressure is high enough to reduce \Tcf , even in good pressure conditions. It has also to be mentioned that the transition broadening is in part related to the fact that above $ \sim $ 4~GPa, \Tcf\textit{(P)} line is becoming steeper with pressure. The $ \Delta $\Tc resulting from fixed experimental uncertainties will consequently increase. In the present case then, it is useful to evaluate the pressure evolution of the fit parameters $ \rho (T)= \rho _{0} + A T^{n} $ obtained at very low temperature to find further evidences for a quantum critical point. 

As we already showed, it was found to be acceptable to fit only down to 1.8~K ($ ^{4} $He cryostat temperatures) at least to get a qualitative behavior. The results from the low temperature fits $ \rho (T)= \rho _{0} + A T^{n} $ presented in figure \ref{LTfit} are consistent with a presence of pressure induced quantum critical point. A sharp peak is observed in the \textit{A(P)} graph and the \textit{n(P)} graph drops sharply to $n \sim $1 as the critical pressure is approached. At low pressures, the \textit{n} exponent is above 2 as expected in the magnetic phase for a Kondo lattice system, and it tends to increase with the magnetic transition temperature. It then decreases until the critical pressure. At that point, \textit{n} is around 1.35-1.4, when measured in a $^{3} $He cryostat. This value is very close to 4/3, given by the spin fluctuation model in the case of a two-dimensional ferromagnet. However the lowest temperature obtained to determine \textit{n} was 0.35~K, which might be too high when close to the critical pressure. \textit{n} is then probably a little underestimated (from our estimations, \textit{n} tends to increase when the temperature decreases in this pressure range). The \textit{A} and $ \rho _{0} $ parameters appear to be less sensitive to the fit temperature range. $\rho _{0} $ slightly increases while approaching $P_{c} $, and then present a stronger decrease. However, its behavior is much different when pressure is applied along the \textit{c}-axis and current along the \textit{b}-axis, probably because of the resistivity anisotropy. It is interesting to notice that at higher pressures, far enough from the critical pressure,  $\rho _{0} $ is even lower than at ambient pressure.

No superconductivity was observed in this compound down to the lowest temperature of 0.35~K reached in this work. This may be due to the ferromagnetic order, as no superconductivity was found in any other Ce-based ferromagnetic compounds such as CeNiSb$_{3}$\cite{Sidorov05Ce} or CeAgSb$_{2}$\cite{Sidorov03,Nakashima03}, presenting many similarities.  Antiferromagnetic order is indeed known to be more propitious to superconductivity, compared to a ferromagnetism \cite{Mathur98}, and it has been shown that \textit{d}-wave singlet pairing in nearly antiferromagnetic metals is generally much stronger than \textit{p}-wave triplet pairing in nearly ferromagnetic metals \cite{Monthoux99}. On another hand, at least four U-based ferromagnetic compounds, which are all Ising-type ferromagnets, were found to be superconductors, at ambient pressure for URhGe \cite{Aoki01} and UCoGe \cite{Huy07}, or under pressure for UGe$_{2}$ \cite{Saxena00} and UIr \cite{Akazawa04}. The residual resistivity ratio of \CVS is low (below 10 over the whole pressure range) with a rather high residual resistivity, above 10 \mWf ,  compared to other superconducting Ce compounds. This may evidence a too strong scattering for the occurrence of exotic superconductivity. As an example, the residual resistivity should not be higher than a few \mW in the case of CePd$_{2}$Si$_{2}$ and CeIn$_{3}$ to observe superconductivity \cite{Mathur98}. On the other hand, for the ferromagnet CeAgSb$_{2}$, no superconductivity was observed in high quality samples with $\rho_{0}$ below 0.5~\mW \cite{Sidorov03,Nakashima03}.
Finally, the lowest temperatures reached of 0.35~K might also be too high to observe any eventual superconductivity.

\section{Conclusion}
We determined the pressure-temperature phase diagram of the ferromagnetic compound \CVSf. An initial increase of \Tc with pressure up to 4.5~GPa (for hydrostatic pressure medium) is observed, followed by the transition being progressively suppressed with further increase of pressure, in agreement with the Doniach model. From the extrapolation of \Tc to zero and the low temperature fits of the resistivity, we find a quantum critical point around 7~GPa. No superconductivity was observed down to 0.35~K. We took advantage of the uniaxial component in the modified Bridgman anvils cell and applied successively the pressure along the three axis. Discrepancies were noticed in the \Tc \textit{(P)} behavior when this slight uniaxial component is applied. The \textit{c}-axis seems to be stiff enough not be sensitive to the uniaxial component, and present a behavior in agreement with pressure conditions closer to hydrostaticity.

Whereas the modified Bridgman cell filled with Fluorinert was not suitable by itself to perform this study, it was shown to be very useful to evaluate the anisotropy in the uniaxial pressure dependence of the crystal. The use of 1:1 n-pentane:isopentane brought a strong improvement in pressure conditions and we are currently working to be able to consistently use it up to 8~GPa with the modified Bridgman cell.

\begin{acknowledgments}
This work was performed in part at Ames Laboratory, US DOE, under contract $\#$ DE-AC02-07CH11358 (E. Colombier, E.~D. Mun, S. L. Bud'ko, and Paul C. Canfield). This project has been supported by the French ANR programs DELICE and CORMAT (G. Knebel, and B. Salce). Part of this work was carried out at the Iowa State University and supported by the AFOSR-MURI grant $ \# $FA9550-09-1-0603 (X. Lin, and P.~C. Canfield). S. L. Bud'ko was also partially supported by the State of Iowa through the Iowa State University.
We would also like to acknowledge Stella Kim for her assistance with pressure cells measurements.
\end{acknowledgments}

\end{document}